\documentclass[aps,prd,twocolumn,nofootinbib,superscriptaddress,showpacs]{revtex4}

\usepackage{amssymb,amsmath}

\usepackage{natbib}

\def\text#1{\mbox{\scriptsize{#1}}}

\def\Oh{\mathcal{O}}

\def\ce{\mbox{ce}}
\def\se{\mbox{se}}

\def\Mc#1{\mbox{Mc}^{(#1)}}
\def\Ms#1{\mbox{Ms}^{(#1)}}

\DeclareMathOperator{\besselj}{J}

\DeclareMathOperator{\mathieuce}{ce}
\DeclareMathOperator{\mathieuse}{se}

\DeclareMathOperator{\mathieumc}{Mc}
\DeclareMathOperator{\mathieums}{Ms}

\DeclareMathOperator{\betafunction}{B}
\DeclareMathOperator{\capitaleta}{H}

\DeclareMathOperator{\jacobisn}{sn}
\DeclareMathOperator{\elliptick}{K}

\begin{document}

\title{Perturbative zero-point energy for a cylinder of elliptical section}
\author{Adrian R. Kitson}
\affiliation{Institute of Fundamental Sciences, Massey University, Private Bag 11 222, Palmerston North, New Zealand}
\author{August Romeo}
\affiliation{Societat Catalana de F{\'\i}sica, Barcelona, Spain}

\date{\today}

\begin{abstract}
We examine the Casimir effect for a perfectly conducting cylinder of elliptical section, taking as reference the known case of circular section. The zero-point energy of this system is evaluated by the mode summation method, using the ellipticity as a perturbation parameter. Mathieu function techniques are applied.
\end{abstract}


\maketitle

\section{Introduction}

Vacuum fluctuations of quantum fields caused by the presence of boundaries produce changes in the zero-point energy, which give rise to the Casimir effect.  Interest in quantum vacuum manifestations, including this phenomenon, has been propelled by new theoretical and experimental advances~\cite{M1,M2,BMM}. Hypothetical prospects of technological applications make even more desirable the knowledge of fundamental aspects of the theory, such as the value of the vacuum energy or its dependence on any of the problem conditions (even the sign of the effect is hard to predict, although for interactions between dielectric bodies some progress has been made in Ref.~\cite{KK}).

Particularly striking is the modification of the zero-point energy caused by a change in boundary shape, even at an infinitesimal level. As Casimir energies prove to be very sensitive to purely geometrical modifications, the subject deserves further consideration. This question has already been addressed for a boundary departing from spherical, which has implications for QCD flux tube models~\cite{KS}.

On the other hand, cylindrical boundaries of circular section have been object of attention under a variety of settings: the perfectly conducting case~\cite{DRM,GR,LNB}, dielectric media with or without light-velocity conservation~\cite{KNN,NLS,KR,BP,CPM1,CPM2,RM1,RM2,Sch}, dispersion~\cite{BN,B,BR}, semitransparent boundary~\cite{CPMK}, varying Robin conditions~\cite{RS}, finite temperature\footnote{For a general discussion about the classical limit in temperature-dependent systems see Ref.~\cite{FMR}.}~\cite{BNP}, coaxial surfaces~\cite{ST}, cosmic strings~\cite{BBST}, etc.  However, deviations from circular section constitute a largely uncharted land. In the present work, we make an incursion into this territory by considering a perfectly conducting and infinitely long cylindrical surface of elliptical section, which slightly deviates from circular shape\footnote{Yet, there are
works~\cite{GO,G} where the elliptical coordinates from scattering theory play a role in the study of General Relativity (averaged) null energy conditions for a Casimir system.}.

In Sec.~2 the solutions to the Maxwell equations in terms of Mathieu functions are considered, and the boundary conditions which determine the eigenmode set are established. The summation of these eigenmodes is studied in Sec.~3, thereby obtaining the zero-point energy of the electromagnetic field as a perturbative expansion in the ellipticity. Sec.~4 offers, as an alternative view, the spectrum modification on the basis of a conformal transformation relating elliptical and circular cases. The conclusions follow in Sec.~5. Relevant properties and results about Mathieu functions have been included in an appendix. Natural units (\(\hbar=c=1\)) are used throughout.

\section{Solutions to Maxwell's equations with perfect conductor boundary conditions}

We will consider the classic problem: find the zero-point energy of an electromagnetic field due to a perfectly conducting, neutral surface \({\mathcal S}\). In the absence of any charge or current density and with time dependence given by \(\exp\left(-i \omega t\right)\), Maxwell's equations in Heaviside-Lorentz units are
\begin{align}
{\bf \nabla}\cdot{\bf E}\left({\bf x}\right)&=0,\\
{\bf \nabla}\cdot{\bf B}\left({\bf x}\right)&=0,\\
{\bf \nabla}\times{\bf E}\left({\bf x}\right)&=i\omega\,{\bf B}\left({\bf x}\right),\\
{\bf \nabla}\times{\bf B}\left({\bf x}\right)&=-i\omega\,{\bf E}\left({\bf x}\right),
\end{align}
where \({\bf E}\left({\bf x}\right)\) is the spatial part of the electric field and \({\bf B}\left({\bf x}\right)\) is the spatial part of the magnetic field. Both vector fields satisfy the vector Helmholtz equation, that is,
\begin{equation}
\left({\bf \nabla}^{2}+\omega^{2}\right){\bf A}\left({\bf x}\right)={\bf 0},
\label{vhe}
\end{equation}
where \({\bf A}\left({\bf x}\right)\) can be \({\bf E}\left({\bf x}\right)\) or \({\bf B}\left({\bf x}\right)\). The vector Laplacian is
\begin{equation}
{\bf \nabla}^{2}{\bf A}\left({\bf x}\right)={\bf \nabla}\,{\bf \nabla}\cdot{\bf A}\left({\bf x}\right)-{\bf \nabla}\times{\bf \nabla}\times{\bf A}\left({\bf x}\right).
\end{equation}
The boundary conditions are
\begin{align}
\left.{\bf \hat{n}}\times{\bf E}\left({\bf x}\right)\right|_{{\bf x}\in{\mathcal S}}&={\bf 0},\label{embc1}\\
\left.{\bf \hat{n}}\cdot{\bf B}\left({\bf x}\right)\right|_{{\bf x}\in{\mathcal S}}&=0,\label{embc2}
\end{align}
where \({\bf \hat{n}}\) is the outward-pointing, unit normal vector to the surface \({\mathcal S}\). In particular, we take $\mathcal{S}$ to be an infinitely long cylinder, along the $z$-axis, with an elliptical section given (in Cartesian coordinates) by the equation
\begin{equation}
\label{ellipse}
\frac{x^2}{a^2}+\frac{y^2}{b^2}=1,
\end{equation}
where \(a\) is the semimajor axis and \(b\) is the semiminor axis. We introduce elliptic cylindrical coordinates, \((\xi, \eta, z)\), a left-handed coordinate system related to Cartesian coordinates by
\begin{align}
x&=f\cosh\left(\xi\right)\cos\left(\eta\right),\\
y&=f\sinh\left(\xi\right)\sin\left(\eta\right),
\end{align}
where \(0\le\xi<\infty\), \(0\le\eta<2\pi\) and \(f\) is the focal length given by
\begin{equation}
f=\sqrt{a^{2}-b^{2}}.
\end{equation}
The ellipticity is defined to be
\begin{equation}
\label{fa}
e=\frac{f}{a}.
\end{equation}
To avoid confusion with the exponential function, its full form, \(\exp\), will be used. In the elliptic cylindrical coordinate system the ellipse~(\ref{ellipse}) adopts the form
\begin{equation}
\xi_{0}^{\mbox{}}=\cosh^{-1}\left(1/e\right).
\end{equation}
The boundary conditions become
\begin{align}
E_{\eta}^{\mbox{}}\left(\xi=\xi_{0}^{\mbox{}}\right)&=0,\label{bcee}\\
E_{z}^{\mbox{}}\left(\xi=\xi_{0}^{\mbox{}}\right)&=0,\label{bcez}\\
B_{\xi}^{\mbox{}}\left(\xi=\xi_{0}^{\mbox{}}\right)&=0.
\end{align}
While the vector Helmholtz equation cannot be solved by separation of variables in all the orthogonal coordinate systems the scalar Helmholtz equation can, the elliptic cylindrical coordinate system is one of the cases where it can~\cite{MF}. Furthermore, since the media inside and outside \({\mathcal S}\) are the same, and thus the speed of light is equal on both sides; we may split the solutions into transverse electric (TE) and transverse magnetic (TM) modes~\cite{CPM1,CPM2,RM1,RM2}. Each vector component can be written in terms of Mathieu and modified Mathieu functions.

\begin{widetext}

\subsection{Interior field}
\label{int}

\subsubsection{TM modes (\(B_{z}^{\mbox{}}=0\))}

It is enough to start considering the solution for one of the nonvanishing field components, which we choose to be \(E_{z}^{\mbox{}}\). The general form will be~\cite{ML}:
\begin{equation}
\label{ez}
E_{z}^{\mbox{}}\left(\xi,\eta,z\right)=
\begin{cases}
{\displaystyle
\sum_{n=0}^{\infty}\,\int \frac{dk}{2\pi}\,C_{n}^{\mbox{}}\left(k\right)\,\Mc{1}_{n}\left(\xi,q\right)\ce_{n}^{\mbox{}}\left(\eta,q\right)\exp\left(ikz\right),}\\
{\displaystyle
\sum_{n=1}^{\infty}\,\int \frac{dk}{2\pi}\,S_{n}^{\mbox{}}\left(k\right)\,\Ms{1}_{n}\left(\xi,q\right)\se_{n}^{\mbox{}}\left(\eta,q\right)\exp\left(ikz\right),}
\end{cases}
\end{equation}
\end{widetext}
where (\(\Mc{1}_{n}\) and \(\Ms{1}_{n}\)) \(\ce_{n}^{\mbox{}}\) and \(\se_{n}^{\mbox{}}\) are the even and odd (modified) Mathieu functions of the first kind, respectively. Notice that there is no zeroth order odd Mathieu function. The coefficients \(C_{n}^{\mbox{}}\left(k\right)\) and \(S_{n}^{\mbox{}}\left(k\right)\) are constants. Equation~(\ref{bcez}) implies that
\begin{align}
\Mc{1}_{n}\left(\xi_{0}^{\mbox{}},q_{n,p}^{{\rm I},{\rm TM}}\right)&=0,\\
\Ms{1}_{n}\left(\xi_{0}^{\mbox{}},\tilde{q}_{n,p}^{{\rm I},{\rm TM}}\right)&=0,
\end{align}
where the superscripts refer to the interior region and TM mode, respectively. The second subscript indexes the \(p\)th zero. The tilde distinguishes the odd eigenmodes from the even ones. The dependence of the eigenmodes on the ellipticity has been suppressed. The eigenfrequencies are obtained through the relation
\begin{equation}
\label{ef}
\omega=\sqrt{\frac{4\,q}{f^{2}}+k^{2}}.
\end{equation}
The other two boundary conditions are automatically satisfied because \(E_{\eta}^{\mbox{}}\) and \(H_{\xi}^{\mbox{}}\) involve the same vanishing factors as in \(E_{z}^{\mbox{}}\)~\cite{ML}.

\subsubsection{TE modes (\(E_{z}^{\mbox{}}=0\))}

The TE mode is obtained by the principle of duality; \(B_{z}^{\mbox{}}\) will have the same form as equation~(\ref{ez}). Since
\begin{equation}
E_{\eta}\propto\frac{\partial}{\partial\xi}B_{z}^{\mbox{}},
\end{equation}
equation~(\ref{bcee}) implies that
\begin{equation}
\left.\frac{\partial}{\partial\xi}B_{z}^{\mbox{}}\right|_{\xi=\xi_{0}^{\mbox{}}}=0.
\end{equation}
Therefore,
\begin{align}
\Mc{1}_{n}\mbox{}'\left(\xi_{0}^{\mbox{}},q_{n,p}^{{\rm I},{\rm TE}}\right)&=0,\\
\Ms{1}_{n}\mbox{}'\left(\xi_{0}^{\mbox{}},\tilde{q}_{n,p}^{{\rm I},{\rm TE}}\right)&=0,
\end{align}
where the prime means differentiation with respect to \(\xi\). The other boundary conditions are also satisfied; it follows that \(B_{\xi}^{\mbox{}}\) vanishes on the boundary because \(B_{\xi}^{\mbox{}}\propto E_{\eta}^{\mbox{}}\) and \(E_{z}^{\mbox{}}=0\) by definition of a TE mode.

\subsection{Exterior field}

Although this part of the solution is not discussed in~\cite{ML}, it can be worked out in terms of functions akin to Hankel functions. The reason for doing so is the physical demand that, at large distances, the field components should behave like cylindrical waves. This is easily met by choosing the linear combinations
\begin{align}
\Mc{3}_{n}\left(\xi, q\right)&=\Mc{1}_{n}\left(\xi, q\right)+i\,\Mc{2}_{n}\left(\xi, q\right),\\
\Mc{4}_{n}\left(\xi, q\right)&=\Mc{1}_{n}\left(\xi, q\right)-i\,\Mc{2}_{n}\left(\xi, q\right),
\end{align}
with corresponding relations for the odd modified Mathieu functions.

There are many parallels between modified Mathieu and Bessel functions. As with Bessel functions of the second kind, the modified Mathieu functions of the second kind are not regular in the interior region. This is why they were not included in the solution for the field inside the cylinder, but shall now be used for the field outside. As we want the waves to be outgoing, the modified Mathieu functions of the third kind are selected.

\subsubsection{TM modes (\(B_{z}^{\mbox{}}=0\))}

The \(z\) component of the electric field will have the same form as equation~(\ref{ez}) but with modified Mathieu functions of the third kind. The boundary conditions imply that
\begin{align}
\Mc{3}_{n}\left(\xi_{0}^{\mbox{}},q_{n,p}^{{\rm II},{\rm TM}}\right)&=0,\\
\Ms{3}_{n}\left(\xi_{0}^{\mbox{}},\tilde{q}_{n,p}^{{\rm II},{\rm TM}}\right)&=0,
\end{align}
where the first superscript refers to the exterior region.

\subsubsection{TE modes (\(E_{z}^{\mbox{}}=0\))}

Analogous to the interior TE modes, we find  
\begin{align}
\Mc{3}_{n}\mbox{}'\left(\xi_{0}^{\mbox{}},q_{n,p}^{{\rm II},{\rm TE}}\right)&=0,\\
\Ms{3}_{n}\mbox{}'\left(\xi_{0}^{\mbox{}},\tilde{q}_{n,p}^{{\rm II},{\rm TE}}\right)&=0.
\end{align}

\section{Regularized zero-point energy}
\label{rzpe}

In natural units, the zero-point energy per lateral unit length amounts to half the sum (including integration for \(k\)) of all the eigenfrequencies given by equation~(\ref{ef}). Since such a quantity is divergent, some form of regularization is called for; we use zeta-function regularization.

The even zeta-functions are
\begin{equation}
\zeta_{n}^{{\mathcal A},{\mathcal B}}\left(s\right)=\sum_{p=1}^{\infty}\int_{-\infty}^{\infty}\frac{dk}{2\pi}\,\left[\omega_{n,p}^{{\mathcal A},{\mathcal B}}\right]^{-s},
\end{equation}
where \({\mathcal A}\in\left\{{\rm I},{\rm II}\right\}\) and \({\mathcal B}\in\left\{{\rm TE},{\rm TM}\right\}\). Let
\begin{equation}
\label{qg}
\frac{4\,q}{f^{2}}=\gamma^{2},
\end{equation}
then equation~(\ref{ef}) becomes
\begin{equation}
\omega=\sqrt{\gamma^{2}+k^{2}}.
\end{equation}
Integrating over \(k\) leaves
\begin{equation}
\zeta_{n}^{{\mathcal A},{\mathcal B}}\left(s\right)=\frac{1}{2\pi}\betafunction\left(\frac{1}{2},\frac{s-1}{2}\right)\sum_{p=1}^{\infty}\left[\gamma_{n,p}^{{\mathcal A},{\mathcal B}}\right]^{1-s},
\end{equation}
where \(\betafunction\) is the beta function. Following similar analysis, the odd zeta-functions are
\begin{equation}
\tilde{\zeta}_{n}^{{\mathcal A},{\mathcal B}}\left(s\right)=\frac{1}{2\pi}\betafunction\left(\frac{1}{2},\frac{s-1}{2}\right)\sum_{p=1}^{\infty}\left[\tilde{\gamma}_{n,p}^{{\mathcal A},{\mathcal B}}\right]^{1-s}.
\end{equation}
There is no odd zeta-function for \(n=0\). The regularized zero-point energy per lateral unit length as a function of ellipticity is
\begin{equation}
\label{zpe}
{\mathcal E}_{C}\left(e\right)=\lim_{s\rightarrow -1}\frac{1}{2}\sum_{{\mathcal A},{\mathcal B}}\left[\sum_{n=0}^{\infty}\zeta_{n}^{{\mathcal A},{\mathcal B}}\left(s\right)+\sum_{n=1}^{\infty}\tilde{\zeta}_{n}^{{\mathcal A},{\mathcal B}}\left(s\right)\right],
\end{equation}
where the limit means analytical continuation to \(s=-1\). As we are summing over the interior and exterior regions we expect the analytical continuation to be finite (see~\cite{GR,LR} and references therein). For this reason, neither arbitrary scales nor additional prescriptions have been introduced.

In terms of the dimensionless variable \(z=a\gamma\), the even zeta-functions with \({\mathcal A}={\rm I}\) and \({\mathcal B}={\rm TM}\) are
\begin{equation}
\label{ezf}
\zeta_{n}^{{\rm I},{\rm TM}}\left(s\right)=a^{s-1}\frac{1}{2\pi}\betafunction\left(\frac{1}{2},\frac{s-1}{2}\right)\sum_{p=1}^{\infty}\left[z_{n,p}^{{\rm I},{\rm TM}}\right]^{1-s}.
\end{equation}
The summation over \(p\) can be written as a contour integral with the help of the argument principle (see e.g.~\cite{schram,R1,R2,R3,LR}) as follows
\begin{widetext} 
\begin{equation}
\label{ci}
\sum_{p=1}^{\infty}\left[z_{n,p}^{{\rm I},{\rm TM}}\right]^{1-s}=\frac{s-1}{2\pi i}\int_{\partial\Omega}dz\,z^{-s}\ln\left[\Mc{1}\left(\xi_{0}^{\mbox{}},q\right)\right],
\end{equation}
where, using equations~(\ref{fa}) and~(\ref{qg}),
\begin{equation}
q=\frac{z^{2}e^{2}}{4}.
\end{equation}
The integration circuit is the boundary of a region \(\Omega\) of the complex \(z\)-plane which contains all the wanted zeros and avoids the origin. Other \({\mathcal A}\) and \({\mathcal B}\) and the odd counterparts follow in a similar fashion.

\subsection{Ellipticity expansion}

The modified Mathieu functions appearing in the contour integrals can be expanded for small ellipticity in a formal series (see Appendix~\ref{mmf}). Equation~(\ref{ci}) becomes
\begin{equation}
\sum_{p=1}^{\infty}\left[z_{n,p}^{{\rm I},{\rm TM}}\right]^{1-s}=\frac{s-1}{2\pi i}\int_{\partial\Omega}dz\,z^{-s}\left[\ln\left(\besselj_{n}^{\mbox{}}\left(z\right)\right)-\frac{z}{4\besselj_{n}^{\mbox{}}\left(z\right)}\left(\besselj_{n}'\left(z\right)-\frac{\delta_{n1}}{2}\besselj_{0}^{\mbox{}}\left(z\right)\right)e^{2}+\cdots\right].
\end{equation}
The corresponding odd sum is
\begin{equation}
\sum_{p=1}^{\infty}\left[\tilde{z}_{n,p}^{{\rm I},{\rm TM}}\right]^{1-s}=\frac{s-1}{2\pi i}\int_{\partial\Omega}dz\,z^{-s}\left[\ln\left(\besselj_{n}^{\mbox{}}\left(z\right)\right)-\frac{z}{4\besselj_{n}^{\mbox{}}\left(z\right)}\left(\besselj_{n}'\left(z\right)+\frac{\delta_{n1}}{2}\besselj_{0}^{\mbox{}}\left(z\right)\right)e^{2}+\cdots\right],
\end{equation}
which is identical (up to \(\Oh\left(e^{4}\right)\)) except for the sign in front of the Kronecker delta. If the even and odd sums are added, then
\begin{equation}
\label{eo}
\sum_{p=1}^{\infty}\left[z_{n,p}^{{\rm I},{\rm TM}}\right]^{1-s}+\sum_{p=1}^{\infty}\left[\tilde{z}_{n,p}^{{\rm I},{\rm TM}}\right]^{1-s}=d\left(n\right)\frac{s-1}{2\pi i}\int_{\partial\Omega}dz\,z^{-s}\left[\ln\left(\besselj_{n}^{\mbox{}}\left(z\right)\right)-\frac{z\besselj_{n}'\left(z\right)}{4\besselj_{n}^{\mbox{}}\left(z\right)}\,e^{2}+\cdots\right],
\end{equation}
where to take into account that there is no odd sum for \(n=0\),
\begin{equation}
d\left(n\right)=
\begin{cases}
1,&n=0,\\
2,&n\ge 1.
\end{cases}
\end{equation}
Integrating termwise (formally), the second term is proportional to the first since
\begin{equation}
\int_{\partial\Omega}dz\,z^{-s}\frac{z\besselj_{n}'\left(z\right)}{\besselj_{n}^{\mbox{}}\left(z\right)}=\left(s-1\right)\int_{\partial\Omega}dz\,z^{-s}\ln\left(\besselj_{n}^{\mbox{}}\left(z\right)\right).
\end{equation}
Thus, equation~(\ref{eo}) simplifies to
\begin{equation}
\sum_{p=1}^{\infty}\left[z_{n,p}^{{\rm I},{\rm TM}}\right]^{1-s}+\sum_{p=1}^{\infty}\left[\tilde{z}_{n,p}^{{\rm I},{\rm TM}}\right]^{1-s}=d\left(n\right)\frac{s-1}{2\pi i}\int_{\partial\Omega}dz\,z^{-s}\ln\left(\besselj_{n}^{\mbox{}}\left(z\right)\right)\left[1-\frac{s-1}{4}\,e^{2}+\cdots\right].
\end{equation}
Using equation~(\ref{ezf}) and its odd counterpart,
\begin{align}
\label{sc}
\sum_{n=0}^{\infty}\zeta_{n}^{{\rm I},{\rm TM}}\left(s\right)+\sum_{n=1}^{\infty}\tilde{\zeta}_{n}^{{\rm I},{\rm TM}}\left(s\right)&=a^{s-1}\frac{1}{2\pi}\betafunction\left(\frac{1}{2},\frac{s-1}{2}\right)\nonumber\\
&\qquad\mbox{}\times\sum_{n=0}^{\infty}d\left(n\right)\frac{s-1}{2\pi i}\int_{\partial\Omega}dz\,z^{-s}\ln\left(\besselj_{n}^{\mbox{}}\left(z\right)\right)\left[1-\frac{s-1}{4}\,e^{2}+\cdots\right].
\end{align}
\end{widetext}
The prefactor is exactly what one gets in the circular cylindrical situation. The other combinations of \({\mathcal A}\) and \({\mathcal B}\) all have the same form as equation~(\ref{sc}), but with their corresponding circular cylindrical prefactors. Thus, equation~(\ref{zpe}) gives
\begin{equation}
\label{ezpe}
{\mathcal  E}_{C}\left(e\right)={\mathcal E}_{C}\left(0\right)\left[1+\frac{1}{2}\,e^{2}+\Oh\left(e^{4}\right)\right],
\end{equation}
where \({\mathcal E}_{C}\left(0\right)\) is the regularized zero-point energy of per lateral unit length of a circular cylinder of radius \(a\). While it is doubtful that equation~(\ref{ezpe}) converges for \(0\le e<1\), at worst it is an asymptotic series as \(e\rightarrow 0\). In either case we may write the next term as \(\Oh\left(e^{4}\right)\).

The numerical value of \({\mathcal E}_{C}\left(0\right)\) is~\cite{DRM,GR,LNB}
\begin{equation}
{\mathcal E}_{C}\left(0\right)\approx -\frac{0.01356}{a^{2}},
\end{equation}
which, together with equation~(\ref{ezpe}) gives
\begin{equation}
{\mathcal E}_{C}\left(e\right)\approx -\frac{0.01356}{a^{2}}\left[1+\frac{1}{2}\,e^{2}+\Oh\left(e^{4}\right)\right],
\end{equation}
where \(a\) is the semimajor axis. Let \(R=\left(a+b\right)/2\), then
\begin{equation}
a=\frac{2\,R}{1+\sqrt{1-e^{2}}},
\end{equation}
and
\begin{equation}
\label{mainresult}
{\mathcal E}_{C}\left(e\right)\approx -\frac{0.01356}{R^{2}}\left[1+\Oh\left(e^{4}\right)\right].
\end{equation}
That is, the zero-point energy per lateral unit length in terms of \(R\) is the same as that for a circular cylinder with radius \(R\), up to quartic corrections in the ellipticity\footnote{This is not unique. For instance, \(R=\sqrt{\left(a^{2}+b^{2}\right)/2}\) would cause the same effect as \(R=\left(a+b\right)/2\), since they only differ at \(\Oh\left(e^{4}\right)\).}. Equation~(\ref{mainresult}) can be related to the existence of a conformal mapping in the complex plane which transforms the ellipse into a circle. This subject is discussed in the next section.

\section{Conformal transformation}

\subsection{Formulation}

Some of the ideas in Refs.~\cite{BRk,By,Kv} suggest the use of an adequate conformal map. We shall employ a transformation taking the interior of the ellipse~(\ref{ellipse}) to the interior of the circle \(w=R\exp\left(i\varphi\right)\), where
\begin{equation}
\label{defR}
R=\frac{a+b}{2}.
\end{equation}
The required map \(w:{\mathbb C}\rightarrow{\mathbb C}\) is~\cite{K}
\begin{equation}
\label{cnfmpp}
w\left(z\right)=R\sqrt{k}\jacobisn\left[\frac{2\elliptick\left(k\right)}{\pi}\sin^{-1}\left(\frac{z}{\sqrt{a^{2}-b^{2}}}\right),k\right],
\end{equation}
where \(\jacobisn\) is a Jacobi elliptic function, \(\elliptick\) is the complete elliptic integral of the first kind and \(k\) depends on the semiaxes through theta functions as follows:
\begin{equation}
k=\left(\frac{\vartheta_{2}^{\mbox{}}\left(0,q\right)}{\vartheta_{3}^{\mbox{}}\left(0,q\right)}\right)^{2},
\end{equation}
where
\begin{equation}
q=\left(\frac{a-b}{a+b}\right)^{2}.
\end{equation}
The \(k\) and \(q\) variables used here should not be confused with those in other sections. The theta functions are given by
\begin{align}
\vartheta_{2}^{\mbox{}}\left(z,q\right)&=2q^{1/4}\sum_{n=0}^{\infty}q^{n\left(n+1\right)}\cos\left(\left(2n+1\right)z\right),\\
\vartheta_{3}^{\mbox{}}\left(z,q\right)&=1+2\sum_{n=1}^{\infty}q^{n^{2}}\cos\left(2n z\right).
\end{align}
In terms of \(e\) and \(R\),
\begin{equation}
q=\left(\frac{a e}{2R}\right)^{4}.
\end{equation}
Using the definitions of the theta functions, for small ellipticity\footnote{In the opposite case one may wish to consider, e.g., $a\to\infty$ and fixed $b$. Using theta function properties, for this limit, \(w(z)=R \tanh\left(\pi z/\left(4b\right)\right)\), which is the transformation taking the region between the parallel lines \(y= \pm ib\) to the interior of the circle. Of course, these lines can be viewed as the 2-D projection of parallel plates.},
\begin{equation}
k=\frac{a^{2}}{R^{2}}e^{2}+\Oh\left(e^{6}\right).
\end{equation}
With the help of
\begin{align}
\elliptick\left(k\right)&=\frac{\pi}{2}+\frac{\pi}{8}k^{2}+\Oh\left(k^{4}\right),\\
\jacobisn\left(u,k\right)&=\sin\left(u\right)\nonumber\\
&\mbox{}-\frac{\left[u-\sin\left(u\right)\cos\left(u\right)\right]\cos\left(u\right)}{4}k^{2}+\Oh\left(k^{4}\right),
\end{align}
the small ellipticity expansion of equation~(\ref{cnfmpp}) is
\begin{equation}
w\left(z\right)=z-\frac{z^{3}}{4R^{2}}e^{2}+\Oh\left(e^{4}\right).
\end{equation}
Therefore,
\begin{equation}
\label{dw}
w'\left(z\right)=1-\frac{3z^{2}}{4R^{2}}e^{2}+\Oh\left(e^{4}\right),
\end{equation}
where the prime denotes differentiation with respect to \(z\). For our purposes, we also need the leading term of the inverse of \(w\), given by \(z\left(w\right)=w+\Oh\left(e^{2}\right)\). With \(w=r\exp\left(i\varphi\right)\), \(z=r\exp\left(i\varphi\right)+\Oh\left(e^{2}\right)\). Using equation~(\ref{dw}),
\begin{equation}
\left|w'\right|^{2}=1-\frac{3r^{2}\cos\left(2\varphi\right)}{2R^{2}}e^{2}+\Oh\left(e^{4}\right).
\end{equation}
Under any conformal mapping, the 2-D Laplacian operator transforms as
\begin{equation}
\label{laptrans}
\nabla_{z}^{2}=\left|w'\right|^{2}\nabla_{w}^{2}.
\end{equation}
We start from the 2-D Hemlholtz equation in the \(w\)-plane
\begin{equation}
\left(\nabla_{w}^{2}+\gamma_{n}^{2}\right){\mathcal U}_{n}^{\mbox{}}=0,
\end{equation}
where the chosen eigenvalue symbol comes from equation~(\ref{qg}). The boundary conditions are set on the image of the ellipse, which is \(w=R\exp\left(i\varphi\right)\), \(0\le\varphi<2\pi\). Since the boundary is circular, the eigenfunctions are of the form
\begin{equation}
{\mathcal U}_{n}^{\mbox{}}=f_{n}^{\mbox{}}\left(r\right)\exp\left(i n\varphi\right),
\end{equation} 
where \(f_{n}^{\mbox{}}\left(r\right)\) is the suitably normalized radial function which satisfies the boundary conditions.

\subsection{Rayleigh-Schr{\"o}dinger expansion}

Next, Dirac notation shall be temporarily adopted in order to take advantage of the Rayleigh-Schr{\"o}dinger method. Supposing that the eigenvalues \(\lambda_{n}^{\mbox{}}\) and eigenstates \(|{\mathcal U}_{n}^{\mbox{}}\rangle\) for a Strum-Liouville problem
\begin{equation}
\label{de}
L |{\mathcal U}_{n}^{\mbox{}}\rangle+\lambda_{n}^{\mbox{}}|{\mathcal U}_{n}^{\mbox{}}\rangle=0
\end{equation}
with given boundary conditions are known, one wonders which are the new eigenvalues \(\overline{\lambda}_{n}^{\mbox{}}\) and eigenstates \(|\overline{\mathcal U}_{n}^{\mbox{}}\rangle\) for the `perturbed problem'
\begin{equation}
\label{depert}
L |\overline{\mathcal U}_{n}^{\mbox{}}\rangle+\overline{\lambda}_{n}^{\mbox{}}|\overline{\mathcal U}_{n}^{\mbox{}}\rangle-\varepsilon_{\text{pert}}^{\mbox{}}r_{\text{pert}}^{\mbox{}}|\overline{\mathcal U}_{n}^{\mbox{}}\rangle=0
\end{equation}
under the same type of boundary condition, where \(\varepsilon_{\text{pert}}^{\mbox{}}\) is some small parameter. Including just modifications to first order in \(\varepsilon_{\text{pert}}^{\mbox{}}\), we make the ans{\"a}tze
\begin{align}
|\overline{\mathcal U}_{n}^{\mbox{}}\rangle&=|{\mathcal U}_{n}^{\mbox{}}\rangle+\varepsilon_{\text{pert}}^{\mbox{}}|{\mathcal V}_{n}^{\mbox{}}\rangle+\Oh\left(\varepsilon_{\text{pert}}^{2}\right),\\
\overline{\lambda}_{n}^{\mbox{}}&=\lambda_{n}^{\mbox{}}+\varepsilon_{\text{pert}}^{\mbox{}}\mu_{n}^{\mbox{}}+\Oh\left(\varepsilon_{\text{pert}}^{2}\right),
\end{align}
and replace them into equation~(\ref{depert}). At \(\Oh\left(\varepsilon_{\text{pert}}^{0}\right)\) equation~(\ref{de}) is recovered, while the \(\Oh\left(\varepsilon_{\text{pert}}^{1}\right)\) contribution yields
\begin{equation}
\left(L+\lambda_{n}^{\mbox{}}I\right)|{\mathcal V}_{n}^{\mbox{}}\rangle=\left(r_{\text{pert}}^{\mbox{}}-\mu_{n}^{\mbox{}}\right)|{\mathcal U}_{n}^{\mbox{}}\rangle,
\end{equation}
where \(I\) is the identity operator. After applying \(\langle{\mathcal U}_{n}^{\mbox{}}|\) on both sides, taking into account the adjoint of equation~(\ref{de}) (with \(L^{\dagger}=L\) understood), we obtain
\begin{align}
\mu_{n}^{\mbox{}}&=\langle{\mathcal U}_{n}^{\mbox{}}| r_{\text{pert}}^{\mbox{}}|{\mathcal U}_{n}^{\mbox{}}\rangle,\\
&=\int\!dg\,r_{\text{pert}}^{\mbox{}}\,{\mathcal U}_{n}^{\ast}{\mathcal U}_{n}^{\mbox{}},\label{int2}
\end{align}
where \(dg\) denotes the integration measure making the \(\left\{{\mathcal U}_{n}^{\mbox{}}\right\}\) orthonormal.

For our studied case, in the notation of equation~(\ref{de}),
\begin{align}
L&=\nabla^{2},\label{L}\\
\lambda_{n}^{\mbox{}}&=\gamma_{n}^{2},\label{lamb}
\end{align}
and the integration in equation~(\ref{int2}) will be on the type
\begin{equation}
\int\!dg=\int_{0}^{2\pi}\!d\varphi\int_{{\mathcal R}}dr\,r,
\end{equation}
where \({\mathcal R}\) is the radial range. Now, application of the inverse of transformation~(\ref{cnfmpp}) to equation~(\ref{de}), and using equation~(\ref{laptrans}), give
\begin{equation}
\label{dez}
\nabla_{z}^{2}\overline{\mathcal U}_{n}^{\mbox{}}+\overline{\gamma}_{n}^{2}|w'|^{2}\overline{\mathcal U}_{n}^{\mbox{}}=0.
\end{equation}
Since the small parameter in the pertubed equation~(\ref{depert}) is identified as
\begin{equation}
\label{pertp}
\varepsilon_{\text{pert}}^{\mbox{}}=e^{2},
\end{equation}
comparison of equations~(\ref{depert}) and~(\ref{dez}) leads to \(\overline{\gamma}_{n}^{2}|w'|^{2}=\overline{\lambda}_{n}^{\mbox{}}-\varepsilon_{\text{pert}}^{\mbox{}}r_{\text{pert}}^{\mbox{}}\), which, taking into account equations~(\ref{dw}),~(\ref{L}),~(\ref{lamb}) and~(\ref{pertp}), yields
\begin{equation}
r_{\text{pert}}^{\mbox{}}=\overline{\gamma}_{n}^{2}\frac{3r^{2}\cos\left(2\varphi\right)}{2R^{2}}.
\end{equation}
With this \(r_{\text{pert}}^{\mbox{}}\), we calculate the \(\Oh\left(\varepsilon_{\text{pert}}^{2}\right)\) contribution to the \(n\)th eigenvalue using equation~(\ref{int2})
\begin{align}
\mu_{n}^{\mbox{}}&\propto\int_{0}^{2\pi}d\varphi\,\cos\left(2\varphi\right),\\
&=0.
\end{align}
Hence, we conclude that, up to \(\Oh\left(e^{4}\right)\), the interior eigenvalues will not change, and neither will the zero-point energy.

\section{Conclusions}

The main result in this work is~(\ref{mainresult}), which shows that the Casimir energy per lateral unit length for an elliptical cylinder has the same value, up to quartic corrections in ellipticity, as for a circular cylinder with radius equal to the mean of the two semiaxes. A quadratic correction appears if the same energy is expressed in terms of one of the semiaxes.

This can be envisaged from a conformal transformation~(\ref{cnfmpp}) which maps the ellipse onto the circle in question and, perturbatively speaking, yields no quadratic contribution.  Such an infinitesimal symmetry preserves the Casimir energy if deformations from circular to elliptical sections (or vice versa) do not go beyond \(\Oh\left(e^{2}\right)\).

\begin{widetext}
Furthermore, there is a significant connection with Kvitsinsky's work~\cite{Kv}. The 2-D elliptical zeta-function is 
\begin{equation}
\zeta^{\text{2-D}}\left(\sigma,e\right)=a^{\sigma}\sum_{{\mathcal A},{\mathcal B}}\left[\sum_{p=1}^{\infty}\left[z_{n,p}^{{\mathcal A},{\mathcal B}}\right]^{-\sigma}+\sum_{p=1}^{\infty}\left[\tilde{z}_{n,p}^{{\mathcal A},{\mathcal B}}\right]^{-\sigma}\right],
\end{equation}
where \(\sigma=s-1\). Following a similar analysis as in Sec.~\ref{rzpe},
\begin{equation}
\zeta^{\text{2-D}}\left(\sigma,e\right)=\zeta^{\text{2-D}}\left(\sigma,0\right)\left[1-\frac{\sigma}{4}e^{2}+\Oh\left(e^{4}\right)\right],
\end{equation}
where \(\zeta^{\text{2-D}}\left(\sigma,0\right)\) is the 2-D zeta-function for a circular boundary of radius \(a\). If \(a=1+e^{2}/2\) and \(b=1\), then
\begin{equation}
\zeta^{\text{2-D}}\left(\sigma,e\right)=\zeta^{\text{2-D}}\left(\sigma,0\right)\left[1+\frac{\sigma}{4}e^{2}+\Oh\left(e^{4}\right)\right],
\end{equation}
where the 2-D circular zeta-function now has unit radius. Comparison of the different notations leads to \(e^{2}\Leftrightarrow 2\alpha\), \(\sigma\Leftrightarrow 2p\), \(\zeta^{\text{2-D}}\left(\sigma,e\right)\Leftrightarrow\zeta\left(p;\mbox{ellipse}\right)\) and \(\zeta^{\text{2-D}}\left(\sigma,0\right)\Leftrightarrow\zeta\left(p;D\right)\), where the objects on the right are those in Ref.~\cite{Kv}. Then, our result coincides with the unnumbered formula below equation~(23) of the referred paper, originally derived for \(p=2,3,\ldots\) and just field modes like in our `\({\rm I},{\rm TM}\)' subset, suggesting that the expression in question could be valid beyond its initial settings.

From the viewpoint of Mathieu functions, one may argue that the obtained relation has been established by virtue of a variable change~(\ref{defR}) in formulas~(\ref{MconemMsonemee}). Unfortunately, the next order in ellipticity is significantly more complicated.

\appendix

\section{Mathieu functions}

\subsection{Mathieu functions}
\label{mfn}

Mathieu's differential equation in canonical form is
\begin{equation}
\label{mde}
\left[\frac{\partial^{2}}{\partial\eta^{2}}+a\left(q\right)-2q\cos\left(2\eta\right)\right]\capitaleta\left(\eta\right)=0.
\end{equation}
Requiring the solutions to be periodic leads to characteristic values for \(a\left(q\right)\). Periodic solutions that are even with respect to \(\eta\) have characteristic values \(a_{n}^{\mbox{}}\left(q\right)\), whereas those that are odd have characteristic values \(b_{n}^{\mbox{}}\left(q\right)\). If \(q=0\), then  \(a_{n}^{\mbox{}}\left(0\right)=b_{n}^{\mbox{}}\left(0\right)=n^{2}\) and the even and odd solutions are the cosine and sine functions. The even and odd (periodic) Mathieu functions are, respectively
\begin{align}
\mathieuce_{n}\left(\eta,q\right)&=\sum_{k=-\infty}^{\infty} c_{n,k}^{\mbox{}}\left(q\right)\cos\left[\left(n+2k\right)\eta\right],\quad n=0,1,2,\ldots,\label{ce}\\
\mathieuse_{n}\left(\eta,q\right)&=\sum_{k=-\infty}^{\infty} \tilde{c}_{n,k}^{\mbox{}}\left(q\right)\sin\left[\left(n+2k\right)\eta\right],\quad n=1,2,3,\ldots.\label{se}
\end{align}
\end{widetext}
They are normalized such that
\begin{equation}
\int_{0}^{2\pi}d\eta\left[\mathieuce_{n}^{\mbox{}}\left(\eta,q\right)\right]^{2}=\int_{0}^{2\pi}d\eta\left[\mathieuse_{n}^{\mbox{}}\left(\eta,q\right)\right]^{2}=\pi.
\end{equation}
The normalized coefficients can be expanded in formal power series. For \(n\ge 2\), the even coefficients are
\begin{align}
c_{n,0}^{\mbox{}}\left(q\right)&=1+\Oh\left(q^{2}\right),\\
c_{n,\mbox{}\pm 1}^{\mbox{}}\left(q\right)&=\mbox{}\mp\frac{1}{4\left(n\pm1\right)}\,q+\Oh\left(q^{2}\right).
\end{align}
For \(n=1\), we have
\begin{align}
c_{1,0}^{\mbox{}}\left(q\right)&=1+\Oh\left(q^{2}\right),\\
c_{1,\mbox{}+1}^{\mbox{}}\left(q\right)&=\mbox{}-\frac{1}{8}\,q+\Oh\left(q^{2}\right).
\end{align}
and for \(n=0\)
\begin{align}
c_{0,0}^{\mbox{}}\left(q\right)&=\frac{1}{\sqrt{2}}+\Oh\left(q^{2}\right),\\
c_{0,\mbox{}+1}^{\mbox{}}\left(q\right)&=\mbox{}-\frac{1}{2\sqrt{2}}\,q+\Oh\left(q^{2}\right),
\end{align}
All other coefficients are \(\Oh\left(q^{2}\right)\) or higher. The odd coefficients are not the same as the even; however, they do agree up to \(\Oh\left(q^{2}\right)\) with one exception:
\begin{equation}
\tilde{c}_{2,\mbox{}-1}^{\mbox{}}\left(q\right)=\Oh\left(q^{2}\right).
\end{equation}

\begin{widetext}

\subsection{Modified Mathieu functions}
\label{mmf}

The modified Mathieu differential equation is
\begin{equation}
\label{mmde}
\left[\frac{\partial^{2}}{\partial\xi^{2}}-a\left(q\right)+2q\cosh\left(2\xi\right)\right]\Xi\left(\xi\right)=0,
\end{equation}
Let \(\zeta=2\sqrt{q}\cosh\left(\xi\right)\), then
\begin{equation}
\label{mmde2}
\left[\left(\zeta^{2}-4q\right)\frac{\partial^{2}}{\partial\zeta^{2}}+\zeta\frac{\partial}{\partial\zeta}-a\left(q\right)-2q+\zeta^{2}\right]\Xi\left(\zeta\right)=0.
\end{equation}
When \(q=0\), \(a\left(0\right)=a_{n}^{\mbox{}}\left(0\right)=b_{n}^{\mbox{}}\left(0\right)=n^{2}\) and equation~(\ref{mmde2}) reduces to the Bessel differential equation. The even and odd modified Mathieu functions of the first kind are, respectively
\begin{align}
\mathieumc_{n}^{\left(1\right)}\left(\xi, q\right)&=\sum_{k=-\infty}^{\infty}d_{n,k}^{\mbox{}}\left(q\right)\besselj_{n+2k}\left(2\sqrt{q}\cosh\left(\xi\right)\right),\qquad n=0,1,2,\ldots,\label{mc1}\\
\mathieums_{n}^{\left(1\right)}\left(\xi, q\right)&=\tanh{\left(\xi\right)}\sum_{k=-\infty}^{\infty}\tilde{d}_{n,k}^{\mbox{}}\left(q\right)\besselj_{n+2k}\left(2\sqrt{q}\cosh\left(\xi\right)\right),\qquad n=1,2,3,\ldots.\label{ms1}
\end{align}
\end{widetext}
The coefficients are related to those in Section~\ref{mfn}
\begin{align}
d_{n,k}^{\mbox{}}\left(q\right)&=\rho\left(q\right)\left(-1\right)^{k}c_{n,k}^{\mbox{}}\left(q\right),\\
\tilde{d}_{n,k}^{\mbox{}}\left(q\right)&=\tilde{\rho}\left(q\right)\left(-1\right)^{k}\left(n+2k\right)\tilde{c}_{n,k}^{\mbox{}}\left(q\right),
\end{align}
where \(\rho\left(q\right)\) and \(\tilde{\rho}\left(q\right)\) are normalization factors. The modified Mathieu functions of the first kind are normalized to have the same asymptotic form as the Bessel functions
\begin{equation}
\mathieumc_{n}^{\left(1\right)}\left(\xi,q\right)\stackrel{\zeta\rightarrow\infty}{\sim}\mathieums_{n}^{\left(1\right)}\left(\xi,q\right)\stackrel{\zeta\rightarrow\infty}{\sim}\besselj_{n}^{\mbox{}}\left(\zeta\right).
\end{equation}
The normalization factors are therefore
\begin{align}
\frac{1}{\rho\left(q\right)}&=\sum_{k=-\infty}^{\infty}c_{n,k}^{\mbox{}}\left(q\right),\\
\frac{1}{\tilde{\rho}\left(q\right)}&=\sum_{k=-\infty}^{\infty}\left(n+2k\right)\tilde{c}_{n,k}^{\mbox{}}\left(q\right).
\end{align}
The expressions in equation~(\ref{mc1}) and~(\ref{ms1}) are absolutely convergence for \(|\cosh\left(\xi\right) |>1\)~\cite{ML}.

Using the results from Section~\ref{mfn} the coefficients can be expanded in formal power series. The even coefficients for \(n\ge 2\) are
\begin{align}
d_{n,0}^{\mbox{}}\left(q\right)&=1-\frac{1}{2\left(n^{2}-1\right)}\,q+\Oh\left(q^{2}\right),\label{dn0}\\
d_{n,\mbox{}\pm 1}^{\mbox{}}\left(q\right)&=\mbox{}\pm\frac{1}{4\left(n\pm 1\right)}\,q+\Oh\left(q^{2}\right).\label{dnpm1}
\end{align}
For \(n=1\)
\begin{align}
d_{1,0}^{\mbox{}}\left(q\right)&=1+\frac{1}{8}\,q+\Oh\left(q^{2}\right),\label{d10}\\
d_{1,\mbox{}+1}^{\mbox{}}\left(q\right)&=\frac{1}{8}\,q+\Oh\left(q^{2}\right),\label{d1p1}
\end{align}
and for \(n=0\)
\begin{align}
d_{0,0}^{\mbox{}}\left(q\right)&=1+\frac{1}{2}\,q+\Oh\left(q^{2}\right),\label{d00}\\
d_{0,\mbox{}+1}^{\mbox{}}\left(q\right)&=\frac{1}{2}\,q+\Oh\left(q^{2}\right).\label{d0p1}
\end{align}
All the other even coefficients are \(\Oh\left(q^{2}\right)\) or higher. The odd coefficients for \(n\ge 2\) are
\begin{align}
\tilde{d}_{n,0}^{\mbox{}}\left(q\right)&=1+\frac{1}{2\left(n^{2}-1\right)}\,q+\Oh\left(q^{2}\right),\\
\tilde{d}_{n,\mbox{}\pm1}^{\mbox{}}\left(q\right)&=\mbox{}\pm\frac{n\pm 2}{4n\left(n\pm 1\right)}\,q+\Oh\left(q^{2}\right).
\end{align}
For \(n=1\)
\begin{align}
\tilde{d}_{1,0}^{\mbox{}}\left(q\right)&=1+\frac{3}{8}\,q+\Oh\left(q^{2}\right),\\
\tilde{d}_{1,\mbox{}+1}^{\mbox{}}\left(q\right)&=\frac{3}{8}\,q+\Oh\left(q^{2}\right).
\end{align}
All the other odd coefficients are \(\Oh\left(q^{2}\right)\) or higher.

\begin{widetext}

\subsubsection{Ellipticity expansion}

Consider the even modified Mathieu functions of the first kind. With \(\xi_{0}^{\mbox{}}=\cosh^{-1}\left(1/e\right)\) and \(q=z^{2}e^{2}/4\), equation~(\ref{mc1}) is
\begin{equation}
\mathieumc_{n}^{\left(1\right)}\left(\xi_{0}^{\mbox{}},q\right)=\sum_{k=-\infty}^{\infty}d_{n,k}^{\mbox{}}\left(z^{2}e^{2}/4\right)\besselj_{n+2k}\left(z\right).
\end{equation}
For \(n\ge 2\), using equations~(\ref{dn0}) and~(\ref{dnpm1}), the formal series in ellipticity is
\begin{equation}
\label{mcse}
\mathieumc_{n}^{\left(1\right)}\left(\xi_{0}^{\mbox{}},q\right)=\besselj_{n}^{\mbox{}}\left(z\right)+\frac{\left(n-1\right)\besselj_{n+2}^{\mbox{}}\left(z\right)-2\besselj_{n}^{\mbox{}}\left(z\right)-\left(n+1\right)\besselj_{n-2}^{\mbox{}}\left(z\right)}{16\left(n^{2}-1\right)}z^{2}\,e^{2}+\Oh\left(e^{4}\right).
\end{equation}
Using Bessel function recurrence relations, equation~(\ref{mcse}) simplifies to
\begin{equation}
\mathieumc_{n}^{\left(1\right)}\left(\xi_{0}^{\mbox{}},q\right)=\besselj_{n}^{\mbox{}}\left(z\right)-\frac{z\besselj_{n}'\left(z\right)}{4}\,e^{2}+\Oh\left(e^{4}\right).
\end{equation}
Similarly, for \(n=0\), using equations~(\ref{d00}) and~(\ref{d0p1})
\begin{equation}
\mathieumc_{0}^{\left(1\right)}\left(\xi_{0}^{\mbox{}},q\right)=\besselj_{0}^{\mbox{}}\left(z\right)-\frac{z\besselj_{0}'\left(z\right)}{4}\,e^{2}+\Oh\left(e^{4}\right),
\end{equation}
that is, the \(n=0\) case follows the same rule as for \(n\ge 2\). Using equations~(\ref{d10}) and~(\ref{d1p1}), the \(n=1\) case is
\begin{equation}
\mathieumc_{1}^{\left(1\right)}\left(\xi_{0}^{\mbox{}},q\right)=\besselj_{1}^{\mbox{}}\left(z\right)-\frac{z}{4}\left(\besselj_{1}'\left(z\right)-\frac{1}{2}\besselj_{0}\left(z\right)\right)e^{2}+\Oh\left(e^{4}\right).
\end{equation}
Collecting all the results, in general, for \(n\ge 0\)
\begin{equation}
\mathieumc_{n}^{\left(1\right)}\left(\xi_{0}^{\mbox{}},q\right)=\besselj_{n}^{\mbox{}}\left(z\right)-\frac{z}{4}\left(\besselj_{n}'\left(z\right)-\frac{\delta_{n 1}^{\mbox{}}}{2}\besselj_{0}\left(z\right)\right)\,e^{2}+\Oh\left(e^{4}\right).
\end{equation}
By similar analysis, the ellipticity expansion for the odd modified Mathieu functions of the first kind is
\begin{equation}
\label{MconemMsonemee}
\mathieums_{n}^{\left(1\right)}\left(\xi_{0}^{\mbox{}},q\right)=\besselj_{n}^{\mbox{}}\left(z\right)-\frac{z}{4}\left(\besselj_{n}'\left(z\right)+\frac{\delta_{n 1}^{\mbox{}}}{2}\besselj_{0}\left(z\right)\right)\,e^{2}+\Oh\left(e^{4}\right).
\end{equation}
The same procedure is repeated for the modified Mathieu functions of the third kind and for the required derivatives.

\begin{acknowledgments}
The authors are grateful to Kimball A. Milton for valuable observations and comments. A. R. wishes to thank Josep M. Parra for useful conversations. A. K. wishes to thank Tony Signal.
\end{acknowledgments}

\end{widetext}


\begin{thebibliography}{39}
\expandafter\ifx\csname natexlab\endcsname\relax\def\natexlab#1{#1}\fi
\expandafter\ifx\csname bibnamefont\endcsname\relax
  \def\bibnamefont#1{#1}\fi
\expandafter\ifx\csname bibfnamefont\endcsname\relax
  \def\bibfnamefont#1{#1}\fi
\expandafter\ifx\csname citenamefont\endcsname\relax
  \def\citenamefont#1{#1}\fi
\expandafter\ifx\csname url\endcsname\relax
  \def\url#1{\texttt{#1}}\fi
\expandafter\ifx\csname urlprefix\endcsname\relax\def\urlprefix{URL }\fi
\providecommand{\bibinfo}[2]{#2}
\providecommand{\eprint}[2][]{\url{#2}}

\bibitem[{\citenamefont{Milton}(2001)}]{M1}
\bibinfo{author}{\bibfnamefont{K.~A.} \bibnamefont{Milton}},
  \emph{\bibinfo{title}{The Casimir Effect: Physical Manifestations of
  Zero-Point Energy}} (\bibinfo{publisher}{World Scientific},
  \bibinfo{year}{2001}).

\bibitem[{\citenamefont{Milton}(2004)}]{M2}
\bibinfo{author}{\bibfnamefont{K.~A.} \bibnamefont{Milton}},
  \bibinfo{journal}{J. Phys.} \textbf{\bibinfo{volume}{A37}},
  \bibinfo{pages}{R209} (\bibinfo{year}{2004}), \eprint{hep-th/0406024}.

\bibitem[{\citenamefont{Bordag et~al.}(2001)\citenamefont{Bordag, Mohideen, and
  Mostepanenko}}]{BMM}
\bibinfo{author}{\bibfnamefont{M.}~\bibnamefont{Bordag}},
  \bibinfo{author}{\bibfnamefont{U.}~\bibnamefont{Mohideen}}, \bibnamefont{and}
  \bibinfo{author}{\bibfnamefont{V.~M.} \bibnamefont{Mostepanenko}},
  \bibinfo{journal}{Phys. Rept.} \textbf{\bibinfo{volume}{353}},
  \bibinfo{pages}{1} (\bibinfo{year}{2001}), \eprint{quant-ph/0106045}.

\bibitem[{\citenamefont{Kenneth and Klich}(2006)}]{KK}
\bibinfo{author}{\bibfnamefont{O.}~\bibnamefont{Kenneth}} \bibnamefont{and}
  \bibinfo{author}{\bibfnamefont{I.}~\bibnamefont{Klich}}
  (\bibinfo{year}{2006}), \eprint{quant-ph/0601011}.

\bibitem[{\citenamefont{Kitson and Signal}(2006)}]{KS}
\bibinfo{author}{\bibfnamefont{A.~R.} \bibnamefont{Kitson}} \bibnamefont{and}
  \bibinfo{author}{\bibfnamefont{A.~I.} \bibnamefont{Signal}},
  \bibinfo{journal}{J. Phys.} \textbf{\bibinfo{volume}{A39}},
  \bibinfo{pages}{6473} (\bibinfo{year}{2006}), \eprint{hep-th/0511048}.

\bibitem[{\citenamefont{DeRaad and Milton}(1981)}]{DRM}
\bibinfo{author}{\bibfnamefont{J.}~\bibnamefont{DeRaad},
  \bibfnamefont{Lester~L.}} \bibnamefont{and}
  \bibinfo{author}{\bibfnamefont{K.~A.} \bibnamefont{Milton}},
  \bibinfo{journal}{Ann. Phys.} \textbf{\bibinfo{volume}{136}},
  \bibinfo{pages}{229} (\bibinfo{year}{1981}).

\bibitem[{\citenamefont{Gosdzinsky and Romeo}(1998)}]{GR}
\bibinfo{author}{\bibfnamefont{P.}~\bibnamefont{Gosdzinsky}} \bibnamefont{and}
  \bibinfo{author}{\bibfnamefont{A.}~\bibnamefont{Romeo}},
  \bibinfo{journal}{Phys. Lett.} \textbf{\bibinfo{volume}{B441}},
  \bibinfo{pages}{265} (\bibinfo{year}{1998}), \eprint{hep-th/9809199}.

\bibitem[{\citenamefont{Lambiase et~al.}(1999)\citenamefont{Lambiase,
  Nesterenko, and Bordag}}]{LNB}
\bibinfo{author}{\bibfnamefont{G.}~\bibnamefont{Lambiase}},
  \bibinfo{author}{\bibfnamefont{V.~V.} \bibnamefont{Nesterenko}},
  \bibnamefont{and} \bibinfo{author}{\bibfnamefont{M.}~\bibnamefont{Bordag}},
  \bibinfo{journal}{J. Math. Phys.} \textbf{\bibinfo{volume}{40}},
  \bibinfo{pages}{6254} (\bibinfo{year}{1999}), \eprint{hep-th/9812059}.

\bibitem[{\citenamefont{Milton et~al.}(1999)\citenamefont{Milton, Nesterenko,
  and Nesterenko}}]{KNN}
\bibinfo{author}{\bibfnamefont{K.~A.} \bibnamefont{Milton}},
  \bibinfo{author}{\bibfnamefont{A.~V.} \bibnamefont{Nesterenko}},
  \bibnamefont{and} \bibinfo{author}{\bibfnamefont{V.~V.}
  \bibnamefont{Nesterenko}}, \bibinfo{journal}{Phys. Rev.}
  \textbf{\bibinfo{volume}{D59}}, \bibinfo{pages}{105009}
  (\bibinfo{year}{1999}).

\bibitem[{\citenamefont{Nesterenko et~al.}(2004)\citenamefont{Nesterenko,
  Lambiase, and Scarpetta}}]{NLS}
\bibinfo{author}{\bibfnamefont{V.~V.} \bibnamefont{Nesterenko}},
  \bibinfo{author}{\bibfnamefont{G.}~\bibnamefont{Lambiase}}, \bibnamefont{and}
  \bibinfo{author}{\bibfnamefont{G.}~\bibnamefont{Scarpetta}},
  \bibinfo{journal}{Riv. Nuovo Cim.} \textbf{\bibinfo{volume}{27N6}},
  \bibinfo{pages}{1} (\bibinfo{year}{2004}), \eprint{hep-th/0503100}.

\bibitem[{\citenamefont{Klich and Romeo}(2000)}]{KR}
\bibinfo{author}{\bibfnamefont{I.}~\bibnamefont{Klich}} \bibnamefont{and}
  \bibinfo{author}{\bibfnamefont{A.}~\bibnamefont{Romeo}},
  \bibinfo{journal}{Phys. Lett.} \textbf{\bibinfo{volume}{B476}},
  \bibinfo{pages}{369} (\bibinfo{year}{2000}), \eprint{hep-th/9912223}.

\bibitem[{\citenamefont{Bordag and Pirozhenko}(2001)}]{BP}
\bibinfo{author}{\bibfnamefont{M.}~\bibnamefont{Bordag}} \bibnamefont{and}
  \bibinfo{author}{\bibfnamefont{I.~G.} \bibnamefont{Pirozhenko}},
  \bibinfo{journal}{Phys. Rev.} \textbf{\bibinfo{volume}{D64}},
  \bibinfo{pages}{025019} (\bibinfo{year}{2001}), \eprint{hep-th/0102193}.

\bibitem[{\citenamefont{Cavero-Pelaez and Milton}(2005)}]{CPM1}
\bibinfo{author}{\bibfnamefont{I.}~\bibnamefont{Cavero-Pelaez}}
  \bibnamefont{and} \bibinfo{author}{\bibfnamefont{K.~A.}
  \bibnamefont{Milton}}, \bibinfo{journal}{Ann. Phys.}
  \textbf{\bibinfo{volume}{320}}, \bibinfo{pages}{108} (\bibinfo{year}{2005}),
  \eprint{hep-th/0412135}.

\bibitem[{\citenamefont{Cavero-Pelaez and Milton}(2006)}]{CPM2}
\bibinfo{author}{\bibfnamefont{I.}~\bibnamefont{Cavero-Pelaez}}
  \bibnamefont{and} \bibinfo{author}{\bibfnamefont{K.~A.}
  \bibnamefont{Milton}}, \bibinfo{journal}{J. Phys.}
  \textbf{\bibinfo{volume}{A39}}, \bibinfo{pages}{6225} (\bibinfo{year}{2006}),
  \eprint{hep-th/0511171}.

\bibitem[{\citenamefont{Romeo and Milton}(2005)}]{RM1}
\bibinfo{author}{\bibfnamefont{A.}~\bibnamefont{Romeo}} \bibnamefont{and}
  \bibinfo{author}{\bibfnamefont{K.~A.} \bibnamefont{Milton}},
  \bibinfo{journal}{Phys. Lett.} \textbf{\bibinfo{volume}{B621}},
  \bibinfo{pages}{309} (\bibinfo{year}{2005}), \eprint{hep-th/0504207}.

\bibitem[{\citenamefont{Romeo and Milton}(2006)}]{RM2}
\bibinfo{author}{\bibfnamefont{A.}~\bibnamefont{Romeo}} \bibnamefont{and}
  \bibinfo{author}{\bibfnamefont{K.~A.} \bibnamefont{Milton}},
  \bibinfo{journal}{J. Phys.} \textbf{\bibinfo{volume}{A39}},
  \bibinfo{pages}{6703} (\bibinfo{year}{2006}).

\bibitem[{\citenamefont{Schaden}(2006)}]{Sch}
\bibinfo{author}{\bibfnamefont{M.}~\bibnamefont{Schaden}}
  (\bibinfo{year}{2006}), \eprint{hep-th/0604119}.

\bibitem[{\citenamefont{Brevik and Nyland}(1994)}]{BN}
\bibinfo{author}{\bibfnamefont{I.}~\bibnamefont{Brevik}} \bibnamefont{and}
  \bibinfo{author}{\bibfnamefont{G.~H.} \bibnamefont{Nyland}},
  \bibinfo{journal}{Ann. Phys.} \textbf{\bibinfo{volume}{230}},
  \bibinfo{pages}{321} (\bibinfo{year}{1994}).

\bibitem[{\citenamefont{Barton}(2001)}]{B}
\bibinfo{author}{\bibfnamefont{G.}~\bibnamefont{Barton}}, \bibinfo{journal}{J.
  Phys.} \textbf{\bibinfo{volume}{A34}}, \bibinfo{pages}{4083}
  (\bibinfo{year}{2001}).

\bibitem[{\citenamefont{Brevik and Romeo}(2006)}]{BR}
\bibinfo{author}{\bibfnamefont{I.}~\bibnamefont{Brevik}} \bibnamefont{and}
  \bibinfo{author}{\bibfnamefont{A.}~\bibnamefont{Romeo}}
  (\bibinfo{year}{2006}), \eprint{hep-th/0601211}.

\bibitem[{\citenamefont{Cavero-Pelaez et~al.}(2006)\citenamefont{Cavero-Pelaez,
  Milton, and Kirsten}}]{CPMK}
\bibinfo{author}{\bibfnamefont{I.}~\bibnamefont{Cavero-Pelaez}},
  \bibinfo{author}{\bibfnamefont{K.~A.} \bibnamefont{Milton}},
  \bibnamefont{and} \bibinfo{author}{\bibfnamefont{K.}~\bibnamefont{Kirsten}}
  (\bibinfo{year}{2006}), \eprint{hep-th/0607154}.

\bibitem[{\citenamefont{Romeo and Saharian}(2001)}]{RS}
\bibinfo{author}{\bibfnamefont{A.}~\bibnamefont{Romeo}} \bibnamefont{and}
  \bibinfo{author}{\bibfnamefont{A.~A.} \bibnamefont{Saharian}},
  \bibinfo{journal}{Phys. Rev.} \textbf{\bibinfo{volume}{D63}},
  \bibinfo{pages}{105019} (\bibinfo{year}{2001}), \eprint{hep-th/0101155}.

\bibitem[{\citenamefont{Feinberg et~al.}(2001)\citenamefont{Feinberg, Mann, and
  Revzen}}]{FMR}
\bibinfo{author}{\bibfnamefont{J.}~\bibnamefont{Feinberg}},
  \bibinfo{author}{\bibfnamefont{A.}~\bibnamefont{Mann}}, \bibnamefont{and}
  \bibinfo{author}{\bibfnamefont{M.}~\bibnamefont{Revzen}},
  \bibinfo{journal}{Annals Phys.} \textbf{\bibinfo{volume}{288}},
  \bibinfo{pages}{103} (\bibinfo{year}{2001}), \eprint{hep-th/9908149}.

\bibitem[{\citenamefont{Bordag et~al.}(2002)\citenamefont{Bordag, Nesterenko,
  and Pirozhenko}}]{BNP}
\bibinfo{author}{\bibfnamefont{M.}~\bibnamefont{Bordag}},
  \bibinfo{author}{\bibfnamefont{V.~V.} \bibnamefont{Nesterenko}},
  \bibnamefont{and} \bibinfo{author}{\bibfnamefont{I.~G.}
  \bibnamefont{Pirozhenko}}, \bibinfo{journal}{Phys. Rev.}
  \textbf{\bibinfo{volume}{D65}}, \bibinfo{pages}{045011}
  (\bibinfo{year}{2002}).

\bibitem[{\citenamefont{Saharian and Tarloyan}(2006)}]{ST}
\bibinfo{author}{\bibfnamefont{A.~A.} \bibnamefont{Saharian}} \bibnamefont{and}
  \bibinfo{author}{\bibfnamefont{A.~S.} \bibnamefont{Tarloyan}}
  (\bibinfo{year}{2006}), \eprint{hep-th/0603144}.

\bibitem[{\citenamefont{Bezerra~de Mello et~al.}(2006)\citenamefont{Bezerra~de
  Mello, Bezerra, Saharian, and Tarloyan}}]{BBST}
\bibinfo{author}{\bibfnamefont{E.~R.} \bibnamefont{Bezerra~de Mello}},
  \bibinfo{author}{\bibfnamefont{V.~B.} \bibnamefont{Bezerra}},
  \bibinfo{author}{\bibfnamefont{A.~A.} \bibnamefont{Saharian}},
  \bibnamefont{and} \bibinfo{author}{\bibfnamefont{A.~S.}
  \bibnamefont{Tarloyan}} (\bibinfo{year}{2006}), \eprint{hep-th/0605231}.

\bibitem[{\citenamefont{Graham and Olum}(2005)}]{GO}
\bibinfo{author}{\bibfnamefont{N.}~\bibnamefont{Graham}} \bibnamefont{and}
  \bibinfo{author}{\bibfnamefont{K.~D.} \bibnamefont{Olum}},
  \bibinfo{journal}{Phys. Rev.} \textbf{\bibinfo{volume}{D72}},
  \bibinfo{pages}{025013} (\bibinfo{year}{2005}), \eprint{hep-th/0506136}.

\bibitem[{\citenamefont{Graham}(2006)}]{G}
\bibinfo{author}{\bibfnamefont{N.}~\bibnamefont{Graham}}, \bibinfo{journal}{J.
  Phys.} \textbf{\bibinfo{volume}{A39}}, \bibinfo{pages}{6423}
  (\bibinfo{year}{2006}), \eprint{hep-th/0601038}.

\bibitem[{\citenamefont{Morse and Feshbach}(1953)}]{MF}
\bibinfo{author}{\bibfnamefont{P.}~\bibnamefont{Morse}} \bibnamefont{and}
  \bibinfo{author}{\bibfnamefont{H.}~\bibnamefont{Feshbach}},
  \emph{\bibinfo{title}{Methods of Theoretical Physics}}
  (\bibinfo{publisher}{McGraw-Hill}, \bibinfo{address}{New York},
  \bibinfo{year}{1953}).

\bibitem[{\citenamefont{McLachlan}(1947)}]{ML}
\bibinfo{author}{\bibfnamefont{N.~W.} \bibnamefont{McLachlan}},
  \emph{\bibinfo{title}{Theory and Application of Mathieu Functions}}
  (\bibinfo{publisher}{Clarendon Press}, \bibinfo{address}{Oxford},
  \bibinfo{year}{1947}).

\bibitem[{\citenamefont{Leseduarte and Romeo}(1996)}]{LR}
\bibinfo{author}{\bibfnamefont{S.}~\bibnamefont{Leseduarte}} \bibnamefont{and}
  \bibinfo{author}{\bibfnamefont{A.}~\bibnamefont{Romeo}},
  \bibinfo{journal}{Annals Phys.} \textbf{\bibinfo{volume}{250}},
  \bibinfo{pages}{448} (\bibinfo{year}{1996}), \eprint{hep-th/9605022}.

\bibitem[{\citenamefont{van Kampen et~al.}(1968)\citenamefont{van Kampen,
  Nijboer, and Schram}}]{schram}
\bibinfo{author}{\bibfnamefont{N.~G.} \bibnamefont{van Kampen}},
  \bibinfo{author}{\bibfnamefont{B.~R.~A.} \bibnamefont{Nijboer}},
  \bibnamefont{and} \bibinfo{author}{\bibfnamefont{K.}~\bibnamefont{Schram}},
  \bibinfo{journal}{Phys. Lett.} \textbf{\bibinfo{volume}{A27}},
  \bibinfo{pages}{307} (\bibinfo{year}{1968}).

\bibitem[{\citenamefont{Leseduarte and Romeo}(1994)}]{R1}
\bibinfo{author}{\bibfnamefont{S.}~\bibnamefont{Leseduarte}} \bibnamefont{and}
  \bibinfo{author}{\bibfnamefont{A.}~\bibnamefont{Romeo}}, \bibinfo{journal}{J.
  Phys.} \textbf{\bibinfo{volume}{A27}}, \bibinfo{pages}{2483}
  (\bibinfo{year}{1994}).

\bibitem[{\citenamefont{Romeo}(1995)}]{R2}
\bibinfo{author}{\bibfnamefont{A.}~\bibnamefont{Romeo}},
  \bibinfo{journal}{Phys. Rev.} \textbf{\bibinfo{volume}{D52}},
  \bibinfo{pages}{7308} (\bibinfo{year}{1995}).

\bibitem[{\citenamefont{Romeo}(1996)}]{R3}
\bibinfo{author}{\bibfnamefont{A.}~\bibnamefont{Romeo}},
  \bibinfo{journal}{Phys. Rev.} \textbf{\bibinfo{volume}{D53}},
  \bibinfo{pages}{3392} (\bibinfo{year}{1996}).

\bibitem[{\citenamefont{Berry and Robnik}(1986)}]{BRk}
\bibinfo{author}{\bibfnamefont{M.~V.} \bibnamefont{Berry}} \bibnamefont{and}
  \bibinfo{author}{\bibfnamefont{M.}~\bibnamefont{Robnik}},
  \bibinfo{journal}{J. Phys.} \textbf{\bibinfo{volume}{A19}},
  \bibinfo{pages}{649} (\bibinfo{year}{1986}).

\bibitem[{\citenamefont{Berry}(1986)}]{By}
\bibinfo{author}{\bibfnamefont{M.~V.} \bibnamefont{Berry}},
  \bibinfo{journal}{J. Phys.} \textbf{\bibinfo{volume}{A19}},
  \bibinfo{pages}{2281} (\bibinfo{year}{1986}).

\bibitem[{\citenamefont{Kvitsinsky}(1996)}]{Kv}
\bibinfo{author}{\bibfnamefont{A.~A.} \bibnamefont{Kvitsinsky}},
  \bibinfo{journal}{J. Phys.} \textbf{\bibinfo{volume}{A29}},
  \bibinfo{pages}{6379} (\bibinfo{year}{1996}).

\bibitem[{\citenamefont{Kober}(1957)}]{K}
\bibinfo{author}{\bibfnamefont{H.}~\bibnamefont{Kober}},
  \emph{\bibinfo{title}{Dictionary of Conformal Representations}}
  (\bibinfo{publisher}{Dover Publications}, \bibinfo{year}{1957}).

\end{thebibliography}
\end{document}